\documentstyle[12pt,epsfig]{article} 
\begin{document} 
\begin{center} 
\vspace{.2in} 
{\bf {TRITON CLUSTERING IN NEUTRON RICH NUCLEI}} 
\end{center} 
\vspace{.4in} 
\begin{center} 
{\bf S. Afsar Abbas (1) and Farooq H Bhat (2)}\\ 
\vspace{.05in} 
1. Centre for Theoretical Physics, JMI, New Delhi-110025, India\\
2. Physics Department, Islamia College of Science and Commerce, 
Srinagar-190001, India
\\ 
\vspace{.3in} 
email afsar.ctp@jmi.ac.in 
\end{center} 
\vspace{.7in} 
\begin{center} 
{\bf Abstract} 
\end{center} 
\vspace{.3in}

Recently, it has been reported that as one goes from oxygen to fluorine, just
the addition of one more proton, provides extraordinary stability to fluorine
which can bind six more neutrons beyond what oxygen can. It is shown here 
that this surprising stability can be understood if neutron rich nuclei, 
$^{24}O$ and $^{27}F$ are treated as bound states of eight and nine-tritons 
respectively. Also the recently discovered $^{42}Si$ is predicted to have  a
bound state structure of fourteen tritons.  

\newpage

It has been found that the doubly magic nucleus $^{28}O$ is unbound and in fact
the heaviest isotope of oxygen is $^{24}O$ [1,2]. On the other hand adding just
one more proton to oxygen leads to a very neutron rich bound state of 
$^{31}F$ [1,2]. Both these results are puzzling. As stated by Sakurai [2], " 
It is remarkable that six additional neutrons can be bound by moving from
oxygen to fluorine, where Z differs by one. The sudden change in stability 
from oxygen to fluorine indicates an extra push of stability for the very
 neutron-rich fluorine isotope". The reason for this "extra push" is not
 understood. As stated in [2], a consistent description of all these 
effects, in the same theoretical framework, is sadly lacking.\\

  Continuing in the same vein, it turns out that N=28 though remains magic
for $^{42}Si$. But it requires that Z=14 become magic too to provide a
 spherically stable $^{42}Si$ [3]. Thus sudden loss of magicity for N=20 but 
its retention for N=28 (though requiring a new proton magic number at Z=14)
is quite puzzling. However note that the state $^{24}O$ is bound. 
In the above perspective one is  
trying to seek common feature between $^{28}O$ and $^{42}Si$. However, 
actually there is not much common between them, 
as one is unbound and the other bound.\\

Let us change our perspective and seek commonality between $^{24}O$ and 
$^{42}Si$. Both of these are bound - and that is an important common 
feature to be accounted for. The other common feature 
is that both of these can be considered as bound states of clusters of 
tritons - $^{24}O$ of 8-tritons and $^{42}Si$ of 14-tritons.\\

Just as even-even and N=Z nuclei can be treated as clusters of N/2 number of 
alphas , can it be that we can treat nuclei ${^{3Z}_{Z} X_{2Z}}$ as a
bound state of Z number of tritons?\\

Indeed, this is exactly what has been suggested and shown to hold empirically
[4,5]. This picture clearly seems to hold true even for a very neutron-rich
nucleus as heavy as $^{42}Si$.      
\\

Let us use this triton clustering picture to understand the above puzzle of as
to why as one goes from oxygen to fluorine, just the addition of one extra
proton, induces extraordinary stability, thereby allowing for the addition 
of as many as six extra neutrons on top of what oxygen can do.\\

Treating all ${^{3Z}_{Z} X_{2Z}}$ nuclei as being a bound state of Z-number of
tritons (${^{3}_{1} H_{2}}$). So ${^{12}Be}$ is 4-t, ${^{27}F}$ is 9-t etc. 
Viewed in this manner, the relevant degree of freedom is tritons treated as
"elementary" entities. Let us pick out and knock out 
a single triton from this unique bound 
state of tritons. This is clearly a single-triton separation energy defined as\\

\begin{center}
$S_t$=BE(Z,N)-BE(Z-1,N-2)-BE(t)    
\end{center}
\vspace{.3in}

Let us plot $S_t$ as a function of the number of tritons in Fig1. 
The experimental data is from [6]. Here, for example, 
triton number 8 would correspond to the nucleus $^{24}O$ and triton number
5 to $^{15}B$ etc. Some remarkable features emerge in this figure. One clear
feature is even-odd effect in triton numbers. Whenever triton number is even,
the triton separation energy is significantly higher than the adjoining odd
triton numbers. This feature is similar to the odd-even effects seen in one
neutron and one proton separation energies plotted with respect to the neutron
and proton numbers respectively. Therein this is seen as evidence for 
identical particle n-n and p-p pairing in nuclei.\\

However the odd-even effect seen in Fig 1. cannot be attributed 
to identical nucleon n-n or p-p pairing. Here we can  
clearly and unambiguously attribute it to two triton pairing i.e. a t-t
pairing effect in these triton constituent nuclei.\\

Note that the pairing n-n and p-p, necessarily arises from a shell structure,
wherein n-n and p-p are most strongly paired if they are in the same shell. 
This analogy can be carried over to the bound states of triton 
in our example here as well. Two
tritons in the same shell seem to be strongly paired, thereby leading to 
a stronger binding with respect to a single unpaired triton.\\

The next most prominent feature in Fig1. is the highest peak in the separation
energy for $N_t$=8 i.e. for $^{24}O$ and an equally sharp dip for $N_t$=9 
i.e. for $^{27}F$.\\

We know that such drops in one-neutron and one-proton separation energies
when going from one Z/N number to the next one is a signal of magicity 
character of a particular Z/N number. In the context of our discussion 
here, magicity means a much stronger binding for a particular number
of tritons as compared to the adjoining number of tritons.
Hence clearly here $N_t$=8 is a magic number with 
respect to different bound states of tritons. So clearly there exists a shell 
structure of the bound states of tritons and wherein there is a large extra
stability for $N_t$=8 and indicating  magicity for this nucleus\\

Let us treat this triton binding potential to be of Harmonic Oscillator (HO)
kind. In HO potential the magic numbers are 2,8,20,40 etc. For our bound state 
of tritons $N_t$=2 is where the system starts and hence may be justifiably 
treated as magic number. Next $N_t$=8, as indicated above, is indeed a
magic number. Unfortunately the data does not go upto $N_t$=20. 
But clearly as per our model here we predict that $^{60}Ca$, as a bound
state of twenty tritons, would be a magic nucleus.
\\

For the case of magicity in one neutron/proton separation energies,
to avoid the jumps due to the odd-even effect, 
one resorts to the smoother plot of two neutron/proton separation energies and
wherein magicity is indicated by kinks in the plot at appropriate neutron/proton
numbers. So here too we also plot two triton separation energies $S_{2t}$ 
as a function of the number of tritons in Fig 1. 
The kink at $N_t$=8 is most prominent, thereby
justifying the magic character of $N_t$=8.\\

We notice that the behaviour of S2t for the region of number of tritons equal to 12-17
appears to be erratic. Note that for triton numbers 12, 14, 15, 16, 17 
the masses are actually "estimated" and are not experimental [6]. 
Infact, the uniform behaviour of the triton separation energies
for the region of triton numbers 2-11 in Fig 1 gives us confidence 
to state that the estimated masses in [6] 
for the relevant nuclei here, have to be corrected and changed.   

To understand the reason that in going from oxygen to fluorine (i.e. an increase
of only one proton ) as many as six extra neutrons can be bound to fluorine, 
let us look at $^{4}He$. It is an exceptionally bound nucleus and its binding
energy is already saturated. Because of its extraordinary stability and
also as its binding is saturated, it does
not allow extra nucleons to be bound to it. 
In fact the next bound even-even nucleus after $^{4}He$ is actually $^{12}C$. 
So extra-stability and saturation 
translates into lack of interest in interacting 
with other nucleons to form bound structures. 
As per Fig 1., $^{24}O$ is such a
strongly bound system of eight tritons. 
Also the fact that $^{27}F$ is so weakly bound system of tritons, in 
comparison, that we can treat ${24}O$ as a saturated system of bound 
tritons. As such (just like $^{4}He$) it has 
no interest in binding extra neutrons. Hence $^{24}O$ is the highest bound 
isotope of oxygen and also 
as to why the putative doubly magic nucleus $^{28}O$ is
found to be unbound. In contrast $^{27}F$ is a much more weakly bound 
system of 9 tritons. As such it could be willing to add extra neutrons, 
in fact six as per
the detection of the bound state $^{31}F$. Hence it appears that this feature 
of oxygen and fluorine is a result of the reality of triton clustering in the 
neutron-rich isotopes of these nuclei.

\newpage

\begin{figure}
\caption{ One and two triton separation energies as a function of triton number}
\epsfclipon
\epsfxsize=0.99\textwidth
\epsfbox{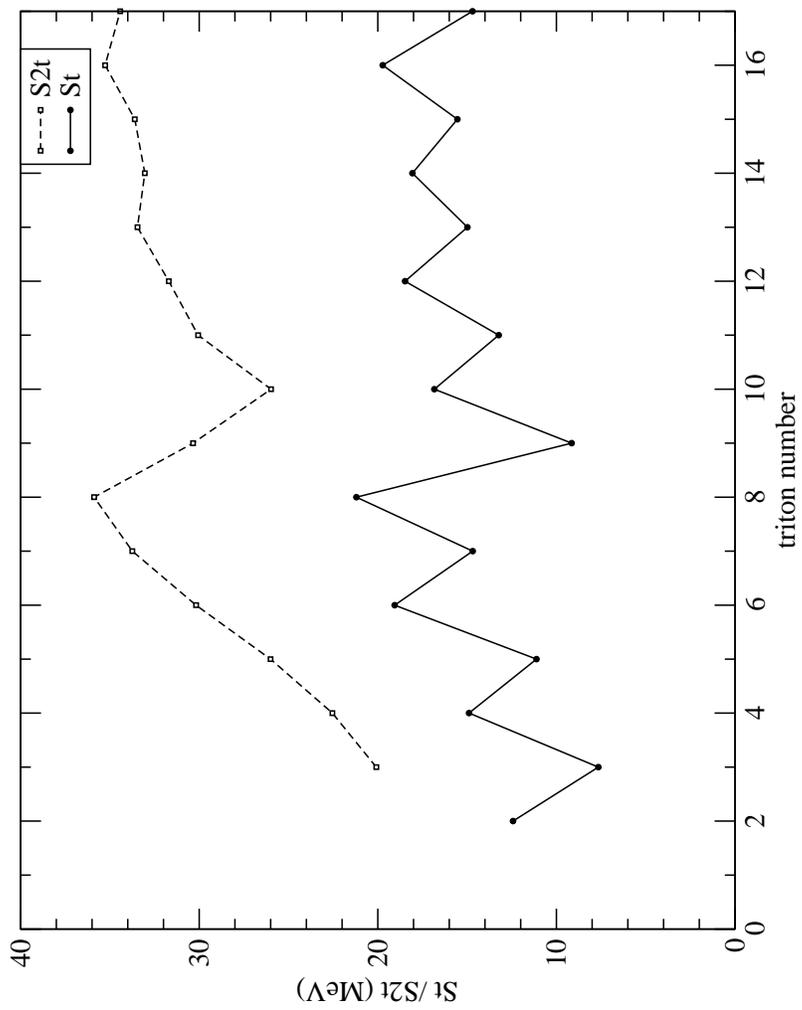}
\end{figure}

\newpage

\vskip .3 cm

\begin{center}
{\bf REFERENCES }
\end{center}
\vspace{.4in}

1. H. Sakurai et al., Phys. Lett {\bf B448} (1994) 180;
\\ Phys. Rev{\bf C54} (1996) R280

\vskip .4 cm

2. H. Sakurai, Eur. Phys. J. {\bf A13} (2002) 49

\vskip .4 cm

3. J. Friedmann et al., Nature {\bf 435} (2005) 922

\vskip .3 cm

4. A. Abbas, Mod. Phys. Lett. {\bf A16} (2001) 755

\vskip .4 cm

5. A. Abbas, Mod. Phys. Lett. {\bf A20} (2005) 2553

\vskip .4 cm

6. G. Audi, A. H. Wapstra and C. Thibault, Nucl. Phys. {\bf A729} 
  (2003) 337

\end{document}